\documentstyle[preprint,aps]{revtex}
\begin{document}

\preprint{{ORNL-CTP-95-06, hep-ph/9510233}}

\title{ Excess Dileptons in High-Energy Nucleus-Nucleus
Collisions}

\author{ Cheuk-Yin Wong$^1$ and Zhong-Qi Wang$^{1,2}$}
\address{ ${}^1$Oak Ridge National Laboratory, Oak Ridge, TN 37831}
\address {\rm and }
\address{ ${}^2$China Institute of Atomic Energy, Beijing, China
}
\date{\today}

\def\sji{\sigma_{{}_{J/\psi}}^{{}^{NN}}}
\def\sabji{\sigma_{{}_{J/\psi}}^{{}^{AB}}}
\def\spaji{\sigma_{{}_{J/\psi}}^{{}^{pA}}}
\def\sabs{\sigma_{\rm abs}}
\def\AAA{{{}_A}}
\def\BB{{{}_B}}
\def\bb{\hbox{\boldmath{$ b $}}}

\maketitle

\begin{abstract}

It has been observed recently that while continuum dilepton production
and open charm production in high-energy $pA$ collisions can be
described in terms of the superposition of $pp$ collisions, dilepton
yields in S+U collisions are in excess of similar extrapolations.
This feature can be explained as arising from the interaction of
gluons produced in different soft baryon-baryon collisions, leading to
additional open-charm pairs in nucleus-nucleus collisions but not in
$pA$ collisions.

\end{abstract}


\narrowtext


High-energy heavy-ion collisions have become the focus of intense
experimental and theoretical research in recent years because of the
possibility of producing a deconfined quark-gluon plasma during such
collisions\cite{QM93,Won94}. Various signals have been proposed to
search for this new phase of quark-gluon matter. In particular,
dilepton production in the environment of a quark-gluon plasma has
been suggested as a way to probe the quark and antiquark momentum
distributions in the plasma \cite{Kaj86}.

Recent experimental measurements on continuum dilepton production in
nucleon-nucleus ($pA$) and nucleus-nucleus ($AB$) collisions at 200
GeV/nucleon \cite{Lou95,Ram95,Mas95,Wur95,Ter95} reveal that dilepton
production in $pA$ collisions in the intermediate energy region
between 1 and 2.5 GeV \cite{Lou95,Ram95,Mas95} can be described by
PQCD (Drell-Yan and open charm) processes as a superposition of $pp$
collisions, using for example the PYTHIA program \cite{Sjo94}.  In the
lower energy region below 1 GeV, the $pA$ data can be described as a
superposition of $pp$ collisions by including contributions from
decaying resonances \cite{Wur95}.  Direct detection of open charm
particles in $pA$ collisions at 800 GeV also indicates that the
production of open charm pairs in $pA$ collisions can be described in
terms of a superposition of $pp$ collisions with essentially no
enhancement \cite{Lei94}.  However, experimental dilepton yields in
$AB$ collisions are in excess of similar $pp$ extrapolations
\cite{Lou95,Ram95,Mas95,Wur95,Ter95}.  These experiments suggest that
there are additional dilepton sources which are present in $AB$
collisions but not in $pA$ collisions
\cite{Lou95,Ram95,Mas95,Wur95,Ter95}.

We focus our attention on dileptons in the intermediate energy region
and study possible additional dilepton sources arising from the
interaction of produced gluons.  In a separate study, Wong recently
examined $\psi'$ and $J/\psi$ suppression in high-energy nuclear
collisions \cite{Won95} where the ratio $\psi'/\psi$ is approximately
a constant in $pA$ collisions, but decreases as the transverse energy
increases in $AB$ collisions \cite{Lou95,Bag95}.  These features can
be explained in terms of a two-component absorption model \cite{Won95}
in which the absorption cross sections for $\psi'$-$N$ and
$J/\psi$-$N$ collisions at high energies are insensitive to the radii
of $c$-$\bar c$ systems in accordance with the Additive Quark Model,
but the cross sections for low energy interactions of $\psi'$ and
$J/\psi$ with produced soft particles are greater for $\psi'$ than for
$J/\psi$ because of the much smaller breakup threshold.  There is
however the ambiguity that these soft particles can be either gluons
or hadrons (as in the comover model \cite{Vog92}).  One way to resolve
the ambiguity is to examine the production of heavy quarks arising
from these soft particles.  While two energetic gluons can interact to
produce a heavy quark pair when the invariant mass of the gluons
exceeds the heavy quark mass threshold, the probability for the
production of a heavy quark pair with two hadrons of the same
invariant mass will be much smaller because the gluons in the hadrons
carry only a fraction of the hadron energies and the fusion of these
gluons does not always have sufficient energy to exceed the heavy
quark mass threshold.  It will be of interest to study the
consequences if the soft particles which disrupt the formation of
$\psi'$ and $J/\psi$ are gluons whose interactions among themselves
lead to the production of charm pairs and additional dileptons.

In soft particle production in a baryon-baryon collision, we envisage
Bjorken's inside-outside cascade picture \cite{Bjo73} or Webber's
picture of gluon branching \cite{Web84} as a $q$ and a $\bar q$ (or a
diquark) pull apart.  After a baryon-baryon soft collision, the
valence quark of one nucleon and the diquark of the other nucleon pull
apart and the gauge field between them is polarized.  Gluons are
emitted at $\tau_{ g}$, and the system hadronizes at $\tau_{\pi}$.
The characteristics of these produced gluons cannot yet be determined
from first principles.  They may acquire an effective mass because of
gluon branching \cite{Web84} and confinement \cite{Won95a}.  Because
hadrons are the products of these gluons, the shape of their momentum
distribution should be close to that of produced hadrons.  Thus,
produced gluons are found predominantly in the central rapidity
region.  Gluons with a large magnitude of rapidity in the
center-of-mass system are also produced with a diminishing
probability.  Before their hadronization, these produced gluons
participate in QCD reactions, much as partons react at high energies.
For example, the reaction of gluons produced in one baryon-baryon
collision with the incipient $c \bar c$ pair produced in another
baryon-baryon collision can lead to the disruption of $\psi'$ and
$J/\psi$ formation in $AB$ collisions \cite{Won95}.  These gluons can
also interact by the gluon fusion mechanism to produce $c \bar c$ and
$s\bar s$ pairs.  Therefore, in the model of produced gluons, the
collision of produced gluons can be additional sources of open charm
and strangeness, accompanied by additional dilepton production.

The space-time trajectories of the baryons and some of the energetic
gluons in a typical $pA$ collision in the nucleon-nucleon
center-of-mass system are depicted schematically in Fig.\ 1$a$.  In
each baryon-baryon collision, gluons are produced after the collision.
To produce an open charm pair will require the collision of energetic
produced gluons moving in opposite directions.  The trajectory of an
energetic right-moving gluon produced in a $pA$ collision will not
cross the trajectory of another energetic left-moving gluon if the
gluon production time $\tau_g$ exceeds a lower limit (which is of the
order of 0.003 fm/c for a beam energy of 200 GeV/nucleon).  We shall
assume that the gluon production time is indeed greater than this
lower limit and the trajectories of energetic gluons traveling in
opposite directions do not cross, as indicated in Fig.\ 1$a$.  Then,
the $gg\rightarrow c \bar c$ reaction for the produced gluons cannot
take place and there is no additional source of open charm production
in $pA$ collisions.  The open charm yield in a $pA$ collision is just
a superposition of open charm yields in $pp$ collisions \cite{Lei94}.
Similarly, the dilepton yield in a $pA$ collision is just a
superposition of dilepton yields in $pp$
collisions\cite{Lou95,Ram95,Mas95}.

In a nucleus-nucleus collision, we can adopt a row-on-row picture and
consider an $AB$ collision as a superposition of the collisions of
rows of projectile nucleons with rows of target nucleons, with various
weights.  We consider specifically a typical one of these rows with a
cross section of the size of the nucleon-nucleon inelastic cross
section, $\sigma_{in}=29.4$ mb.  Within this row, the $N_B\times N_A$
baryon-baryon collisions can be arranged in a space-time diagram as
depicted schematically in Fig.\ 1$b$.  In some of these baryon-baryon
collisions, right-moving energetic gluons can collide with left-moving
energetic gluons, depending on the gluon production time and the
separation of the two collision events.  As a model of additional
source of charm production, we shall assume that the production time
for energetic gluons is less than a certain limit so that energetic
gluons from spatially adjacent baryon-baryon collisions (such as the
collision $A$ and the collision $C$ in Fig.\ 1$b$) will collide.
Then, many other produced gluons can also collide.  For example, an
energetic right-moving gluon produced at $B$ can collide with an
energetic left-moving gluons produced at $C$ and $D$ (Fig.\ 1$b$).  In
this case, as distinctly different from $pA$ collisions, an $AB$
collision will have additional sources of charm production arising
from the collision of energetic gluons produced in separate
baryon-baryon collisions, leading to dilepton production in excess of
$pp$ extrapolations.  The total number of these gluon-gluon collisions
for a given pair of left-moving and right-moving gluons is
\begin{eqnarray}
\label{eq:gc}
N_{\rm col}= N_B(N_B-1)N_A(N_A-1)/4\,,
\end{eqnarray}
which depends on the impact parameter $\bbox{b}$ and the projectile
transverse coordinate $\bbox{b}_B$.  Experimentally, the ratio of the
measured dilepton yield to the dilepton yield calculated on the basis
of PQCD has been measured as a function of the transverse energy,
which can be related to the impact parameter \cite{Lou95}.  The
theoretical PQCD yield for the hard processes (Drell-Yan and open
charm production) at a given impact
parameter $\bbox{b}$ is
\begin{eqnarray}
{d N_{\l^+l^-}^{PQCD} \over dMdy} (M,y;\bbox{b})=
{ d\sigma_{l^+l^-}^{PQCD}(M,y) \over dM dy}
I_{pp}(\bbox{b})
\,,
\end{eqnarray}
where $d\sigma_{l^+l^-}^{PQCD}(M,y)/dM dy$ is the dilepton
differential cross section from the Drell-Yan and the open charm
processes for a $pp$ collision, $I_{pp}(\bbox{b})$ is
\begin{eqnarray}
I_{pp}(\bbox{b})=\int { d \bbox{b}_B \over \sigma_{in}^2} B
T_B(\bbox{b}_B)\sigma_{in} A T_A(\bbox{b}_A) \sigma_{in}\,,
\end{eqnarray}
$T_B(\bbox{b}_B)$ is the thickness function of $B$ \cite{Won94},
$B T_B(\bbox{b}_B) \sigma_{in} AT_A(\bbox{b}_A)\sigma_{in}  $
is the number of baryon-baryon collisions in the row with a cross
section of $\sigma_{in}$ at $\bbox{b}_B$, and
$\bbox{b}_A=\bbox{b}-\bbox{b}_B$.

{}From the collision of produced gluons, there is an additional dilepton
yield in an $AB$ collision at an impact parameter $\bbox{b}$ given by
\begin{eqnarray}
\label{eq:add}
{dN_{\l^+l^-}^{gg} \over dMdy} (M,y;\bbox{b})=
{ d\sigma_{l^+l^-}^{gg}(M,y) \over dM dy}
I_{gg}(\bbox{b})\,,
\end{eqnarray}
where $d\sigma_{l^+l^-}^{gg}(M,y)/dM dy$ is the dilepton differential
cross section from the collision of produced gluons from one
baryon-baryon collision with produced gluons from another
baryon-baryon collision through the intermediate production of $c \bar
c$ pairs.  Using Eq.\ ({\ref{eq:gc}), the quantity $I_{gg}(\bbox{b})$
is
\begin{eqnarray}
I_{gg}(\bbox{b})=\int { d \bbox{b}_B  \over \sigma_{in}^2}
\theta B T_B(\bbox{b}_B)\sigma_{in} [B T_B(\bbox{b}_B)\sigma_{in}-1]
A T_A(\bbox{b}_A) \sigma_{in} [A T_A(\bbox{b}_A) \sigma_{in}-1] /4\,,
\end{eqnarray}
where the step function $\theta=\Theta(A T_\AAA(\bb_\AAA)
\sigma_{in}-1) \Theta (B T_\BB(\bb_\BB) \sigma_{in} -1) \Theta(A-1)
\Theta(B-1)$ with $\Theta (x)=1$ for $x > 0$ and $\Theta (x)=0$ for $x
\le 0$ is to insure that there is no additional source of dileptons in
$pA$ collisions from the collision of produced gluons.  Thus, the ratio
of the total dimuon yield to the PQCD dimuon yield is
\begin{eqnarray}
{dN_{\mu^+\mu^-}({\rm Total~})/dMdy \over dN_{\mu^+\mu^-}( {PQCD} )/dMdy}~
(M,y;\bbox{b})
= 1 + c \, (M,y) { I_{gg}(\bbox{b}) \over  I_{pp}(\bbox{b}) } \,.
\end{eqnarray}
where the quantity  $c(M,y)$
is
\begin{eqnarray}
\label{eq:11}
c(M,y)=
{d\sigma_{\mu^+\mu^-}^{gg}(M,y)/dM dy
\over d\sigma_{\mu^+\mu^-}^{PQCD}(M,y)/dM dy }.
\end{eqnarray}
For a given mass and rapidity window,
the ratio of the total numbers of
$N_{\mu^+\mu^-}$ to the numbers of $N_{\mu^+\mu^-}$ expected from PQCD
is then
\begin{eqnarray}
\label{eq:rat}
{N_{\mu^+\mu^-}({\rm Total~})\over N_{\mu^+\mu^-}( {PQCD} )}~
(\bbox{b})
= 1 + <c \, (M,y)> { I_{gg}(\bbox{b}) \over  I_{pp}(\bbox{b}) } \,,
\end{eqnarray}
where
\begin{eqnarray}
\label{eq:rat2}
<c(M,y)>=\sigma_{\mu^+\mu^-}^{gg}/\sigma_{\mu^+\mu^-}^{PQCD},
\end{eqnarray}
$\sigma_{\mu^+\mu^-}^{gg}$ and $\sigma_{\mu^+\mu^-}^{PQCD}$ are the
corresponding integrated dimuon cross sections within the mass and
rapidity window.

We can estimate the quantity $c(M,y)$ by making a simple description
of the produced gluon momentum distribution.  Hadrons being the
products of the produced gluons, the momentum distribution for the
produced gluons should have the same shape as that of the produced pions.
Using the charged pion distribution given by Eqs.\ (2.4)-(2.8) of
Ref.\ \cite{Won89}, we describe the produced gluon distribution by
\begin{eqnarray}
\label{eq:dis}
{dn \over dy} ({\rm gluon})=A_g [(1-x_+)(1-x_-)]^{a}
\end{eqnarray}
where
\begin{eqnarray}
a= [ 3.5 +0.7 \ln (\sqrt{s}/{\rm GeV})],
\end{eqnarray}
\begin{eqnarray}
x_{\pm}=m_{gT}e^{\pm(y-y_{\pm})}/m_N\,,
\end{eqnarray}
$y_+$ and $y_-$ are the beam and target nucleon rapidity respectively,
$m_{gT}$ and $m_N$ are the gluon transverse mass and nucleon mass
respectively.  The constant $A_g$ can be obtained by requiring that
the total energy carried by produced gluons is the same as the total
energy carried by their hadronized products.  To determine the dimuon
production cross sections from PQCD processes in a $pp$ collision, we
calculate the Drell-Yan cross section in the next-to-leading order
with the CTEQ parton distribution \cite{Lai95}, and the open-charm
cross section in the lowest-order with a $K$-factor equal to 3.0
\cite{Vog92,Ber88}.

For the NA38 measurement with the intermediate mass range (1.5
GeV$<M<2.5$ GeV) and the pseudorapidity range (3.$<\eta<$4.0), the
quantity $<c(M,y)>$ calculated with the gluon distribution of Eq.\
(\ref{eq:dis}) and a gluon transverse mass $m_{gT}= 1 $ GeV is 0.278.
The theoretical (Total/PQCD) ratio as given by Eq.\ (\ref{eq:rat}) is
shown as the solid curve in Fig.\ 2, as a function of the transverse
energy which is related to the impact parameter \cite{Bag90}.  The
theoretical ratio increases as the transverse energy increases.  The
general trend of the theoretical results agrees approxmately with the
preliminary (Mesured/PQCD) data of NA38 \cite{Lou95}.  The NA38 data
of 1990 and 1991 have large errors and a different behavior.  The
HELIOS-3 data \cite{Mas95} cover the same mass region but a different
rapidity range ($3.5<\eta<5.2$).  The value of $<c(M,y)>$ calculated
with the gluon distribution of Eq.\ (\ref{eq:dis}) for the mass and
rapidity region of HELIOS-3 is 0.14.  On the other hand, the HELIOS-3
(Measured/PQCD) data range from 1.74 to 2.51, with large statistical
errors and the data can be described by Eq.\ (\ref{eq:rat}) with
$<c(M,y)> = 0.85 $.  Thus, the HELIOS-3 data are at variance with the
present theoretical results and the naive extrapolation of the NA38
data.  It is important to refine the measurements and the analyses so
as to reduce the experimental uncertainties and differences.  It is
encouraging that there is partial agreement of the present model with
the 1992 data of NA38, but further confirmation is needed.

The present model of dilepton excess is based on the concept that the
excess dileptons arise from open charm production in excess of what is
predicted by $pp$ superpositions.  It is important to confirm this
aspect of nucleus-nucleus collisions experimentally by direct open
charm detections.  Gluons being efficient producers of open charm
particles, this presence of excess open charm partciles will suggest
the likely presence of additional gluon sources.  Open charm pairs and
$J/\psi$ particles are produced in the same intermediate processes.
If the produced gluons can be an additional source of open charm
production, they can be also be an additional source of $J/\psi$
production.  If the presence of excess open charm particles is
confirmed by further measurements, $J/\psi$ suppression in
nucleus-nucleus collisions needs to be reanalyzed in light of this
additional contribution from produced gluons.  The collision of
produced gluons will be an additional source of strange quark pairs
and will also enhance strangeness and $\phi$ production in high-energy
nucleus-nucleus collisions, as compared to what one can extrapolate
from $pp$ collisions.  This is qualitatively consistent with the
enhanced strangeness and $\phi$ production in $AB$ collisions observed
experimentally \cite{Roh94,Ram95}.

The presence of the produced gluons, if confirmed, will suggest that
the medium created by heavy-ion collisions is richly endowed with
gluons produced by soft processes.  The abundance of these gluons
makes it favorable to form gluon matter which can be the precursor of
the quark-gluon plasma and can bring the plasma into existence at a
later stage.

\acknowledgements

One of us (Z. Q. Wang) would like to thank Dr.\ M. Strayer and Dr.\
F. Plasil for their hospitality at Oak Ridge National Laboratory.  The
authors would like to thank Dr.\ C. Louren\c co and Dr.\ G. Young for
helpful comments.  This research was supported in part by the Division
of Nuclear Physics, U.S. Department of Energy under Contract
No. DE-AC05-84OR21400 managed by Lockheed Martin Energy Systems.

\begin{figure}[htbp]
\caption
{ Schematic space-time diagram in the nucleon-nucleon center-of-mass
system, with the time axis pointing upward. ($a$) is for a $pA$
collision and ($b$) is for an $AB$ collision.  The trajectories of the
baryons are given as solid lines.  The trajectories of right-moving
energetic gluons produced in some baryon-baryon collisions are shown
as long-dashed lines and the trajectories of left-moving energetic
gluons by short-dashed lines.  Some baryon-baryon collisions are shown
as solid circles and some collisions between energetic gluons are
shown as open circles.}
\label{fig1}
\end{figure}

\begin{figure}[htbp]
\caption{The ratio of the total $\mu^+ \mu^-$ yield in S+U collisions
at 200A GeV to the $\mu^+ \mu^-$ yield calculated from PQCD, as a
function of the transverse energy $E_T^0$.  Data points are from the
NA38 Collaboration \protect\cite{Lou95}.  }
\label{fig2}
\end{figure}

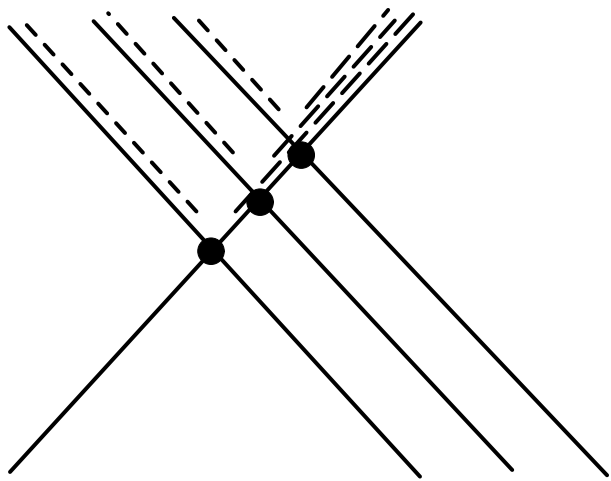
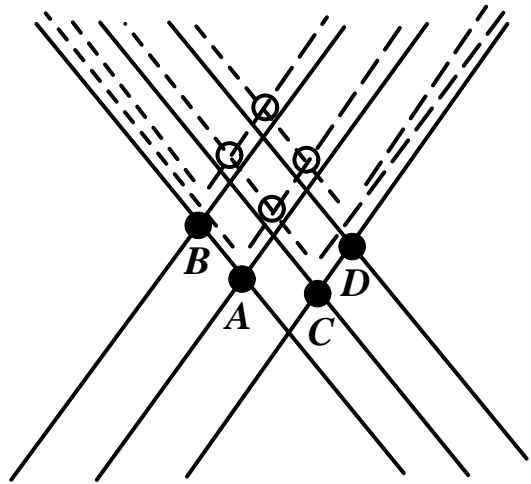
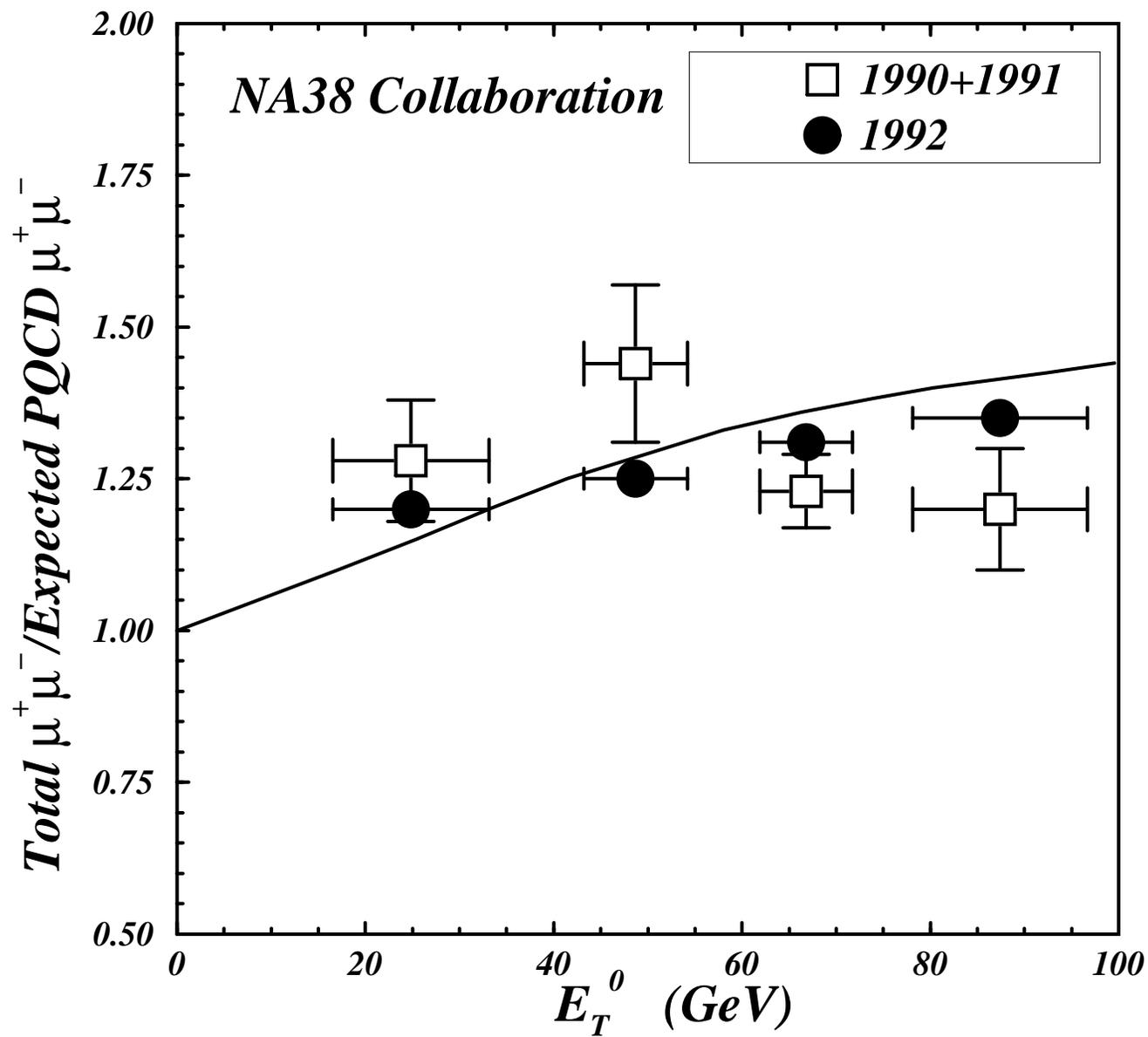
\end{document}